\documentclass[14pt,a4paper,notitlepage]{extarticle}
\usepackage[utf8]{inputenc}
\usepackage[english]{babel}
\usepackage[OT1]{fontenc}
\usepackage{amsmath}
\usepackage{amsfonts}
\usepackage{amssymb}
\usepackage{theorem}
\usepackage[dvips]{graphicx}
\graphicspath{{noiseimages/}}

\usepackage[left=2cm,right=2cm,top=2cm,bottom=2cm]{geometry}

\begin{document}
\date{}

\title{Pseudo-Gauss distribution: \\ a reversible counterpart of the true turbulent velocity gradient distribution\\
\vspace{1cm}
\large{A.V. Kopyev,$^{1,}$\footnote{kopyev@td.lpi.ru} $\;$ K.P. Zybin$^{1,2,}$\footnote{zybin@lpi.ru}}\\
\vspace{0.5cm}
\small{$^1$ \textit{P.N. Lebedev Institute of Physics, Theory Department, 119991 Moscow, Russia}\\
$^2$ \textit{Higher School of Economics, Mathematical Department, 101000 Moscow, Russia}}}

\maketitle
\title{}
\date{}

\abstract{\noindent On the grounds of both widely known
experimental and numerical data of the strain-rate tensor statistical
properties in the fully developed incompressible turbulent flow and
the integral transformations deduced in the article, some important
statistical properties of the flow, known as kinematical, are found
to be dynamical (i.e. taking place purely in a turbulent flow). A
new pseudo-Gauss distribution is introduced in the article and, as a
consequence of these properties, a considerable feature
distinguishing this distribution from the Gauss one is found to be
the general feature of the turbulent statistics. Thus, the new
distribution is proved to be a reversible counterpart of the true
turbulent velocity gradient distribution instead of the Gauss one.}

\section*{1.Introduction}

\noindent The velocity field in the fully developed turbulent flow
is known to be stochastic. Its statistical properties have attracted
the interest of researchers in physics, mechanics and mathematics
for more than a century, but the problem is still far from being
solved. Previously, a new model \cite{Zybin}, which gives a physical
interpretation of the known multifractal model, was formulated. The
large-scale velocity gradients play a significant role in this
model. Now, we try to focus our interest on the statistical
properties of the velocity gradients in the fully-developed
turbulent flow.\vspace{2cm}

\noindent\textit{Full velocity gradient statistical
description}\vspace{0.5cm}

\noindent Velocity gradients compose the tensor, which can be
defined as  value $A_{ij}$  in a given point $\vec{x}_0$ from the
following expansion of the velocity field in $\vec{x}_0$
neighborhood:

\begin{equation}\label{expan}
u_i\left(\vec{x}\right)=u_i\left(\vec{x}_0\right)+A_{ij}\left(\vec{x}_0\right)\left(x_j-{x_0}_j\right)+O\left(|\vec{x}-\vec{x}_0|^2\right)
\end{equation}

\noindent This tensor is called the velocity gradient tensor and as it is evident from (\ref{expan}):

\begin{equation}\label{VGT}
A_{ij}\left(\vec{x}_0\right)=\dfrac{\partial u_i}{\partial x_j}|_{\vec{x}=\vec{x}_0}
\end{equation}

\noindent It follows from the incompressibility of the flow that the trace of the velocity gradient tensor is equal to zero:

\begin{equation}\label{incomp}
A_{ii}=0
\end{equation}

\noindent Under these conditions, the velocity gradient tensor has 8 independent components. It has 5 rotational invariants which are independent of the frame orientation. For instance, the invariants can be chosen as \cite{Meneveau}:

\begin{equation}\label{maininvar}
Q=-\dfrac{1}{2}A_{im}A_{mi} \;\;\;\; R=-\dfrac{1}{3}A_{im}A_{mn}A_{ni}
\end{equation}
\begin{equation}\label{sinvar}
Q_S=-\frac{1}{2}S_{im}S_{mi} \;\;\;\; R_S=-\frac{1}{3}S_{im}S_{mn}S_{ni}
\end{equation}
\begin{equation}\label{vinvar}
V^2=S_{in}S_{im}\omega_m\omega_n
\end{equation}

\noindent Here $S_{ij}$ is a strain-rate tensor and $\vec{\omega}$ - vorticity of the flow. They are both components of the velocity tensor and can be defined by decomposition:

\begin{equation}\label{decomp}
A_{ij}=\frac{1}{2}\left(A_{ij}+A_{ji}\right)+\frac{1}{2}\left(A_{ij}-A_{ji}\right)=S_{ij}-\frac{1}{2}\epsilon_{ijm}\omega_m
\end{equation}

\noindent Here $\epsilon_{ijm}$ is the antisymmetric tensor. Because
of $SO(3)$ symmetry the probability density functions (PDFs) of
velocity gradients depend on rotational invariants only
\cite{Il'yn}. It could be expressed in terms of invariants
(\ref{maininvar})-(\ref{vinvar}) or in any other five independent
invariants, which are functions of invariants
(\ref{maininvar})-(\ref{vinvar}).\vspace{0.5cm}

\noindent\textit{Symmetric velocity gradients statistical description}\vspace{0.5cm}

\noindent This article focuses on the statistical properties of the
symmetrical part of decomposition (\ref{decomp}). One can see from
(\ref{sinvar}) that invariants $Q_S$ and $R_S$ are functions of
strain-rate tensor $S_{ij}$ only. Due to the incompressibility and
symmetrical arguments $S_{ij}$ has only two invariants which are
independent of the frame orientation. Thus, the statistical
properties of strain-rate tensor $S_{ij}$ can be considered
independently in terms of $Q_S$ and $R_S$ invariants or in other two
independent invariants, which must be functions of $Q_S$ and $R_S$.
For instance, new invariants, namely, the maximum
$\left(\lambda_1\right)$ and the minimum $\left(\lambda_3\right)$
eigenvalues of strain-rate tensor $S_{ij}$ could be chosen instead
of $Q_S$ and $R_S$ (due to the incompressibility, the third eigen
value $\lambda_2$, which is greater than $\lambda_3$ but less than
$\lambda_1$, is equal to $-\lambda_1-\lambda_3$). The expressions of
$Q_S$ and $R_S$ in terms of $\lambda_1$ and $\lambda_3$ take the
form:

\begin{equation}\label{Qlam}
Q_S=-\left(\lambda_1^2+\lambda_1\lambda_3+\lambda_3^2\right)
\end{equation}
\begin{equation}\label{Rlam}
R_S=\lambda_1\lambda_3\left(\lambda_1+\lambda_3\right)
\end{equation}

\noindent The inverse transform to (\ref{Qlam}), (\ref{Rlam}) can be shown to be unique, but much more complicated.

\noindent There are other parameters, single PDFs of which do not
give the full statistic information of the strain but are useful in
considering of fluid particle deformation tendencies.

\begin{equation}\label{beta}
\beta=\dfrac{\sqrt{6}\lambda_2}{\sqrt{\lambda_1^2+\lambda_2^2+\lambda_3^2}}
\end{equation}
\begin{equation}\label{s}
s=\dfrac{-3\sqrt{6}\lambda_1\lambda_2\lambda_3}{\left(\lambda_1^2+\lambda_2^2+\lambda_3^2\right)^\frac{3}{2}}
\end{equation}

\noindent They were introduced and analyzed in articles
\cite{Ashurst, LR} respectively. In an incompressible flow both of
them can change in the interval $\left[-1;1\right]$. Each of these
values gives a unique value of eigenvalues ratio
$\lambda_1:\lambda_2:\lambda_3$ but says nothing about their
absolute values. The expression coupling two parameters $\beta$ and
$s$ has been given in the article \cite{LR}, so they cannot be
chosen as two independent parameters of $S_{ij}$ and, thus, it is
surely senseless to find the joint PDF of the
parameters.\vspace{0.5cm}

\noindent\textit{Physical interpretation of velocity gradients}\vspace{0.5cm}

\noindent Note that defined tensors $A_{ij}$ and $S_{ij}$, vorticity
vector $\vec{\omega}$ and the $S_{ij}$ eigen values (with the
corresponding eigen vectors) have a physical interpretation in terms
of fluid particles - macroscopic but small volumes of the moving
flow. Let $\vec{x}_0$ in (\ref{VGT}) be the position vector of the
fluid particle center of mass. If the fluid particle dimensions are
small enough to apply the linear expansion (\ref{expan}), velocity
gradient tensor $A_{ij}$ entirely governs its dynamics in its
center-of-mass system in a given moment: the rotation and the
deformation of the particle are governed by vorticity vector
$\vec{\omega}$ and strain-rate tensor $S_{ij}$ accordingly. As for
the deformation of the fluid particle, three principal axes of the
strain can be distinguished in the directions of $S_{ij}$ eigen
vectors with the exponential rate of strain in these directions.
Namely, the expansion with relative rate $\lambda_1$ occurs in the
direction of the eigen vector corresponding to these directions, the
contraction with rate $|\lambda_3|$ in its eigen vector direction
and either the expansion or the contraction (depending on the sign
of $\lambda_2$) with rate $|\lambda_2|$ take place in the
corresponding eigen vector direction \cite{Loyczansky}. Hence, the
ratio $\lambda_1:\lambda_2:\lambda_3$ (either parameter $\beta$ or
$s$) characterizes the shape of the ellipsoidal deformation of an
initially spherical fluid particle. For example, if $\beta=+1$ (or
$s=+1$) then $\lambda_1:\lambda_2:\lambda_3=1:1:-2$  and the
initially spherical fluid particle becomes a pancake-like. If
$\beta=-1$ (or $s=-1$) then $\lambda_1:\lambda_2:\lambda_3=2:-1:-1$
and it tends to be a filament.\vspace{0.5cm}

\noindent\textit{Generalization of velocity gradients on large scales}\vspace{0.5cm}

\noindent If we consider so large fluid particles that the expansion
(\ref{expan}) becomes wrong, the velocity gradient tensor, defined
in (\ref{VGT}), fails to govern their dynamics. However, it is those
particles, which are known to be of huge interest in the turbulence
theory \cite{Frisch}. Therefore, the definition of the velocity
gradient tensor (in \cite{Xu2} it is called "true" velocity gradient
tensor) should be generalized on different scales, for example, with
the help of the following averaging:

\begin{equation}\label{genVGT}
A_{ij}^R\left(\vec{x}_0\right)=\dfrac{1}{V^R}\int_{V^R}A_{ij}\left(\vec{x}_0+\vec{r}\right)d\vec{r}
\end{equation}

\noindent Where $V^R$ is the spherical volume with the center in the
point $\vec{x}_0$ , which radius is equal to $R$. It is possible to
split generalizing tensor $A_{ij}^R$ into symmetrical and
antisymmetric parts $S_{ij}^R$ and
$-\frac{1}{2}\epsilon_{ijm}\omega_m^R$ analogous to $A_{ij}$
(\ref{decomp}). The traces of tensors $A_{ij}^R$ and $S_{ij}^R$ are
obviously equal to zero, so it has five and two nonzero orientation
invariants respectively, which are defined in the same way as it was
done previously. Therefore, general statistical analysis is
analogous for both true and generalized gradient tensors. Due to
this, the results deduced in the article are valid for both
tensors.\vspace{0.5cm}

\noindent\textit{Short overview of the previous results}\vspace{0.5cm}

\noindent All above discussed quantities were calculated in many
direct numerical simulations (DNS) and were measured in a wide range
of experiments (see the overview \cite{Wallace} and the references
in it). There are also  many phenomenological theoretical models for
these quantities (see the overview \cite{Meneveau} and the
references in it). Note that most of numerical, experimental and
theoretical articles on the velocity gradients deal with joint PDF
of $Q$ and $R$ velocity gradient tensor invariants (\cite{Meneveau,
Wallace} and references).

\noindent Concerning strain-rate tensor invariants, note that DNS
are usually focused on the joint PDF of invariants $Q_S$ and $R_S$
(e.g. \cite{Ooi, Soria}) and, on the contrary, in experiments
researchers usually prefer to measure single distributions of
$\lambda_1$, $\lambda_2$ and $\lambda_3$ (e.g. \cite{Luthi,
Gulitski})\footnote{Sometimes single distributions of $Q_S$
\cite{Soria} and $R_S$ \cite{Luthi, Gulitski} can be found both in
numerical and experimental works.}. Single $\beta$ and $s$ PDFs were
measured and calculated in a considerable amount of various fully
developed incompressible and slightly compressible flows. Their
properties has been investigated since they were calculated with low
Reynolds numbers ($\beta$ \cite{Ashurst, LR, Kerr, She}, $s$
\cite{LR}) and measured by hot-wire probes ($\beta$
\cite{Tsinober}). There are modern precise calculations ($\beta$
\cite{Xu2, Buxton}, $s$ \cite{Buxton}) and measurements by both
hot-wire probes ($\beta$ and $s$ \cite{Buxton}) and PIV methods
($\beta$  \cite{Xu2}). The important property of $\beta$ PDF is that
it vanishes at the end points of its domain (i.e. at points
$\beta=\pm 1$). It can be shown \cite{LR} that this means $s$ PDF
boundedness at $s=\pm 1$, which are end points of its
domain.\vspace{0.5cm}

\noindent\textit{Present article results}\vspace{0.5cm}

\noindent In contradiction to the known results of the article
\cite{LR}\footnote{And quite many subsequent articles that refer to
its conclusions (e.g. overview \cite{Meneveau}, interpretation of
experimental \cite{Tao, Higgins, Kang, Chamecki} and DNS \cite{Kang,
Chamecki} results, phenomenological model \cite{Chevillard} and
development of the ideas of \cite{LR} on the compressible flows
\cite{Chumakov})}, we will show that the properties of single
$\beta$ and $s$ PDFs give non-trivial information about turbulent
statistics (see section 4). Before this, in section 3, we will show
that in case of the Gauss velocity distribution these properties do
not take place (i.e. $\beta$ PDF does not vanish at $\beta=\pm 1$
and $s$ PDF tends to infinity in $s=\pm 1$) and will give the
reversibility analog of the Gauss distribution that we have called
the pseudo-Gauss distribution, for which these properties take
place. All the results will be found by means of the integral
transformations between different PDFs that we will deduce in
section 2. Note that it is convenient to give the resulting property
of turbulent statistics not only in terms of invariants $Q_S$ and
$R_S$ but also via "shuffled" eigenvalues of the strain-rate tensor
and their joint PDF, which we will introduce in section 2.

\noindent In our opinion, the contradiction of our results with the
previous ones is associated with our account of pre-exponential
factor in the joint PDF of invariants $Q_S$ and $R_S$ in Gauss case,
which has not been accounted before (see section 3).

\section*{2.Joint and single PDFs of different strain-rate tensor invariants and relations between them}

\noindent\textit{"Shuffled" and "ordered" eigen values}\vspace{0.5cm}

\noindent Let $\lambda_1$, $\lambda_2$ and $\lambda_3$ be eigen
values of the strain-rate tensor $S_{ij}$ ordered by the condition
$\lambda_1\ge\lambda_2\ge\lambda_3$ as they are defined in the
introduction. Let us introduce the procedure, which we call
"shuffling". This means that we set the "ordered" eigenvalues
$\lambda_1$, $\lambda_2$ and $\lambda_3$ into one-to-one
correspondence with so-called "shuffled" eigenvalues $\mu_1$,
$\mu_2$ and $\mu_3$ letting each  be equal to one of three
$\lambda_i$ with the same probability $\frac{1}{3}$. Due to the
one-to-one correspondence, it is evident that the shuffled
eigenvalues hold the incompressibility condition
$\mu_1+\mu_2+\mu_3=0$, too. Moreover, two of the shuffled eigen
values (e.g. $\mu_1$ and $\mu_2$) can be marked as new strain-rate
tensor invariants like $Q_S$ and $R_S$ or $\lambda_1$ and
$\lambda_3$. Thus, as we have mentioned in introduction, PDFs of the
functions depending on the strain-rate tensor can be expressed in
terms of shuffled eigen values.\vspace{0.5cm}

\noindent\textit{Joint PDFs of the shuffled eigen values ($\mu$ PDFs)}\vspace{0.5cm}

\noindent Let us denote the joint PDF of the shuffled eigen values
$\mu_i$ and $\mu_j$ ($i\ne j$) $f_{ij}\left(x,y\right)$. By the
construction of the procedure we have 4 simple properties for $\mu$
PDFs:

\begin{itemize}

\item[1)] $\mathrm{Domain}\left[f_{ij}\left(x,y\right)\right]=\mathbb{R}^2$

\item[2)] $f_{ij}\left(x,y\right)=f_{ji}\left(x,y\right)$

\item[3)] $f_{12}\left(x,y\right)=f_{23}\left(x,y\right)=f_{13}\left(x,y\right)$

\item[4)] $f_{ij}\left(x,y\right)=f_{ij}\left(y,x\right)=f_{ij}\left(x,-x-y\right)$

\end{itemize}

\noindent These properties make  "shuffled" eigenvalues more
convenient than the ordered ones and we will use them and their
joint PDF hereafter\footnote{The more detailed discussion of the
ordered eigenvalues and their joint PDF see in Appendix 1.}. The
inverse relations between the shuffled eigen values and the ordered
ones are explicit and are given by the formulas:

\begin{equation}\label{lamb1}
\lambda_1=\max\left(\mu_1,\mu_2,-\mu_1-\mu_2\right)
\end{equation}
\begin{equation}\label{lamb3}
\lambda_3=\min\left(\mu_1,\mu_2,-\mu_1-\mu_2\right)
\end{equation}

\noindent Thus, the given procedure is reversible. The four properties of the $\mu$ PDFs can be taken into account by only considering the following functions:

\begin{equation}
\label{phi}
f_{ij}\left(x,y\right)=\phi\left(-\left(x^2+xy+y^2\right),xy(x+y)\right)
\end{equation}

\noindent where $\left(x;y\right)\in\mathbb{R}^2$.\vspace{1.5cm}

\noindent\textit{Joint $Q_S$ and $R_S$ PDF}\vspace{0.5cm}

\noindent Let us find the relation between introduced PDF
$f_{ij}\left(x;y\right)$ and joint $Q_S$ and $R_S$ PDF $f_{Q_S R_S}$
. Invariants $Q_S$ and $R_S$ are expressed by $\mu_i$ and $\mu_j$
for any $i$ and $j$ ($i\ne j$) in the following way:

\begin{equation}\label{QRmu}
\left\{
\begin{aligned}
 Q_S&=-\left(\mu_i^2+\mu_i\mu_j+\mu_j^2\right)\\
 R_S&=\mu_i\mu_j\left(\mu_i+\mu_j\right)
\end{aligned}
\right.
\end{equation}

\noindent One can show that exactly six pairs of $\mu_i$ and $\mu_j$
correspond to one pair of values $Q_S$ and $R_S$.\footnote{As
transform (\ref{Qlam}), (\ref{Rlam}) is unique, nonuniqueness of
reverse transformation, in fact, is associated with "shuffling"}
This divides the space $\left(\mu_i;\mu_j\right)$ in 6 independent
sectors, separated from each other by the lines on which two of the
three eigenvalues are equal:

\begin{equation}\label{mulines}
\left\{
\begin{aligned}
& \mu_i=\mu_j\\
& \mu_i=-\mu_i-\mu_j\\
& \mu_j=-\mu_i-\mu_j
\end{aligned}
\right.
\end{equation}

\noindent The domain of $Q_S$ and $R_S$ ($D_{Q_S R_S}$), which is
shown in Fig.1, is bounded by the curves
$R_S=\pm\frac{2}{3\sqrt{3}}\left(-Q_S\right)^\frac{3}{2}$ in the
lower half plane  \cite{Ooi, Soria}. To find the relation between
the introduced joint PDF and the joint PDF of $Q_S$ and $R_S$
invariants the Jacobian of the transformation (\ref{QRmu}) is
needed. Then:

\begin{equation}\label{fQRmu}
f_{Q_S
R_S}\left(q;r\right)=6\cdot\phi\left(q;r\right)\begin{Vmatrix}
\dfrac{\partial\left(\mu_i,\mu_j\right)}{\partial\left(Q_S,R_S\right)}
\end{Vmatrix}_{Q_S=q;R_S=r}=\dfrac{6\cdot\phi\left(q;r\right)}{\sqrt{-4q^3-27r^2}}
\end{equation}

\noindent Where 6 is the number of $\mu_i$ and $\mu_j$ pairs, which
correspond to the pair of values $Q_S$ and $R_S$ from (\ref{QRmu}).
Also note that $\left(Q_S;R_S\right)\in D_{Q_S R_S}$.

\begin{figure}[h]
\centering
\includegraphics[scale=1]{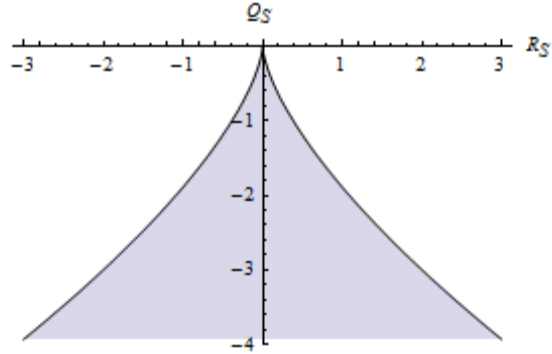} \caption{\small{$D_{Q_S R_S}$: the domain of $Q_S$ and $R_S$ invariants}}
\end{figure}

\noindent Hence, knowing $\mu$ PDF we can find the $Q_S$ and $R_S$
PDF by applying (\ref{fQRmu}). From (\ref{QRmu}) and (\ref{fQRmu})
we get the inverse relation:

\begin{multline}\label{fmuQR}
f_{ij}\left(x;y\right)=\\=\dfrac{|\left(x-y\right)\left(2x+y\right)\left(x+2y\right)|}{6}f_{Q_S R_S}\left(-\left(x^2+xy+y^2\right),xy(x+y)\right)
\end{multline}

\noindent where $\left(x;y\right)\in\mathbb{R}^2$.\vspace{0.5cm}

\noindent\textit{Single $s$ PDF}\vspace{0.5cm}

\noindent Using (\ref{s}), it is possible to get $s$ PDF $f_s$ from $f_{Q_S R_S}$. It can be done by the integral transformation:

\begin{equation}\label{fsQRdeducing}
f_s\left(y\right)=\int_{-\infty}^0
dq\int_{-\frac{2}{3\sqrt{3}}\left(-q\right)^{\frac{3}{2}}}^{\frac{2}{3\sqrt{3}}\left(-q\right)^{\frac{3}{2}}}f_{Q_S
R_S}\left(q;r\right)\delta\left(y-\dfrac{3\sqrt{3}r}{2\left(-q\right)^\frac{3}{2}}\right)dr
\end{equation}

\noindent Here $\delta\left(y-g\left(q;r\right)\right)$ is Dirac
delta function on the surface $g\left(q;r\right)$ \cite{GSh}
convoluting with $f_{Q_S R_S}$ in its domain $D_{Q_S R_S}$, which is
shown in Fig. 1. The calculation (\ref{fsQRdeducing}) could be
reduced to the following expression:

\begin{equation}\label{fsQR}
f_s\left(y\right)=\frac{4}{3\sqrt{3}}
\int_0^{+\infty} f_{Q_S R_S}\left(-t^2;t^3\frac{2y}{3\sqrt{3}}\right)t^4 dt
\end{equation}

\noindent where $s\in\left[-1;1\right]$. From the transformation (\ref{fsQR}) and (\ref{fQRmu}) the similar transformation between the $s$ PDF and the $\mu$ PDF could be determined:

\begin{equation}\label{fsmu}
f_s\left(y\right)=\frac{4}{\sqrt{3}\sqrt{1-y^2}}
\int_0^{+\infty} \phi\left(-t^2;t^3\frac{2y}{3\sqrt{3}}\right)t dt
\end{equation}\vspace{0.5cm}

\noindent\textit{Single $\beta$ PDF}\vspace{0.5cm}

\noindent Now, let us find analogous integral transformations for parameter $\beta$. To do it, let us use the explicit relations between parameters $\beta$ and $s$ given in \cite{LR}:

\begin{equation}\label{sbetas}
\left\{
\begin{aligned}
& \beta=2\sin{\left(\frac{1}{3}\arcsin{s}\right)}\\
& s=\beta\left(3-\beta^2\right)/2
\end{aligned}
\right.
\end{equation}

\noindent Determining the Jacobian of transformation (\ref{sbetas}) and using (\ref{fsQR}) we can find:

\begin{multline}\label{fbetaQR}
f_\beta\left(z\right)=f_s\left(\frac{z\left(3-z^2\right)}{2}\right)
\begin{vmatrix} \dfrac{ds}{d\beta} \end{vmatrix}_{\beta=z}=\\=\frac{2}{\sqrt{3}}\left(1-z^2\right)\int_0^{+\infty} f_{Q_S R_S}\left(-t^2;t^3\frac{z\left(3-z^2\right)}{3\sqrt{3}}\right)t^4 dt
\end{multline}

\noindent where $z\in\left[-1;1\right]$. Analogously we can determine the integral transformation from joint PDF $\phi$ to PDF $f_\beta$:

\begin{equation}\label{fbetamu}
f_\beta\left(z\right)=\frac{4\sqrt{3}}{\sqrt{4-z^2}}\int_0^{+\infty} \phi\left(-t^2;t^3\frac{z\left(3-z^2\right)}{3\sqrt{3}}\right)t dt
\end{equation}

\section*{3. Gauss and pseudo-Gauss distribution}

\noindent\textit{Gauss distribution case}\vspace{0.5cm}

\noindent Let us consider the random velocity field with Gauss
statistics. In the article (\cite{Shtilman} Appendix B) it was shown
that from the Gaussian distribution of the flow velocity follows the
Gaussian distribution of the velocity gradient tensor elements. From
the form given in \cite{Shtilman} one can write for the strain-rate
tensor eigen values PDF:

\begin{equation}\label{Gaussmu}
f_{ij}\left(x;y\right)=\phi\left(-\left(x^2+x y+y^2\right)\right)=\frac{\sqrt{3}}{2\pi M^2}\exp{\left(\dfrac{-x^2-x y-y^2}{M^2}\right)}
\end{equation}

\noindent Here $M$ is a parameter characterizing the distribution
dispersion, which in the incompressible case is governed only by the
mean value of squared vorticity $\omega^2$ \cite{Shtilman}. To find
$Q_S$ and $R_S$ joint PDF let us take into account (\ref{fQRmu}):

\begin{equation}\label{GaussQR}
f_{Q_S R_S}\left(q;r\right)=\dfrac{3\sqrt{3}\exp{\left(q/M^2\right)}}{\pi M^2\sqrt{-4q^3-27r^2}}
\end{equation}

\noindent Substitution (\ref{Gaussmu}) in (\ref{fbetamu}) and (\ref{fsmu}) gives $\beta$ and $s$ PDFs for the Gauss velocity distribution:

\begin{equation}\label{Gaussbeta}
f_\beta\left(z\right)=\dfrac{3}{\pi \sqrt{4-z^2}}
\end{equation}
\begin{equation}\label{Gausss}
f_s\left(y\right)=\dfrac{1}{\pi \sqrt{1-y^2}}
\end{equation}

\noindent These PDFs are shown in Fig. 2. We see that in this case
$f_\beta$ does not vanish at the points $\pm1$ as it was concluded
in \cite{LR} (see figure 1 from \cite{LR}). Moreover, in contrast
with \cite{LR} $f_s$ becomes infinity in the points $\pm1$. In our
opinion, this contradiction is associated with our account of
pre-exponential factor $\left(-4q^3-27r^2\right)^{-1/2}$ in
(\ref{GaussQR}), which was not accounted before. To prove this
statement we now consider $Q_S$ and $R_S$  joint PDF without the
pre-exponential factor.\vspace{0.5cm}

\begin{figure}[h]
\center{\includegraphics[scale=1]{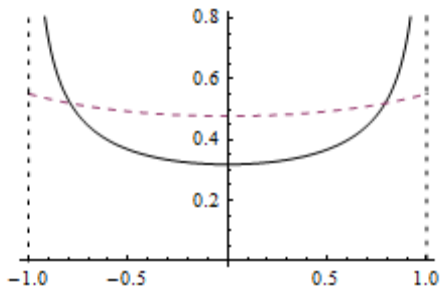}} \caption{\small{PDFs $f_s$
(the full curve with the vertical asymptotes) and $f_\beta$ (the
dotted curve) in case of Gauss velocity field.}}
\end{figure}

\noindent\textit{Pseudo-Gauss distribution case}\vspace{0.5cm}

\noindent Now let us consider the case of symmetric exponential
distribution $f_{Q_S R_S}$ (without the pre-exponential factor) that
we have called a pseudo-Gauss distribution:

\begin{equation}\label{PseudoGaussQR}
f_{Q_S R_S}\left(q;r\right)=\dfrac{\sqrt{3}\exp{\left(q/A^2\right)}}{\sqrt{\pi} A^5}
\end{equation}

\noindent In contrast to $M$ from (\ref{Gaussmu}), parameter $A$
still has no intuitive physically meaningful interpretation because
we do not have the pseudo-Gauss distribution as a consequence of any
specific velocity field like in the previous subsection. The inverse
problem of finding the velocity distribution from distribution
(\ref{PseudoGaussQR}) is not considered in the present article and
the question still remains open.

\noindent By substituting (\ref{PseudoGaussQR}) in (\ref{fmuQR}) we have:

\begin{equation}\label{PseudoGaussmu}
f_{ij}\left(x;y\right)=\dfrac{\sqrt{3}\left|\left(x-y\right)\left(2x+y\right)\left(2y+x\right)\right|}{6 \sqrt{\pi} A^5}\exp{\left(\dfrac{-x^2-x y-y^2}{A^5}\right)}
\end{equation}

\noindent One can see that this $f_{ij}$ has a very specific shape
as it vanishes on the lines $x=y$, $x=-x-y$ and $y=-x-y$, that is,
when two of three strain-rate eigenvalues are equal $\beta=\pm1$ (or
$s=\pm1$). Using (\ref{fsQR}), (\ref{fbetaQR}) and
(\ref{PseudoGaussQR}) after some developments, we have:

\begin{equation}\label{PseudoGaussbeta}
f_\beta\left(z\right)=\frac{3}{4}\left(1-z^2\right)
\end{equation}
\begin{equation}\label{PseudoGausss}
f_s\left(y\right)=\frac{1}{2}
\end{equation}

\noindent Derived PDFs $f_\beta$ and $f_s$ (Fig. 3) are exactly the
same as the ones in figure 1 in the article \cite{LR}.\vspace{0.5cm}

\begin{figure}[h]
\vspace{-5ex}
\begin{center}
\includegraphics[scale=1]{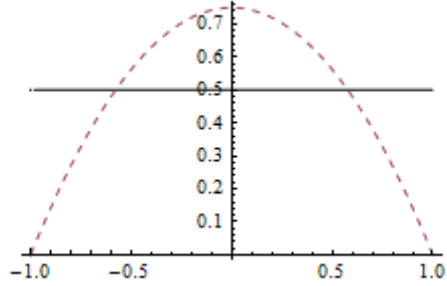}
\end{center}
\caption{\small{PDFs $f_s$ (the full curve) and $f_\beta$ (the
dotted one) in case of pseudo-Gauss velocity field.}}
\end{figure}

\noindent\textit{$Q$-distribution cases}\vspace{0.5cm}

\noindent Let us notice that there is no parameter $M$ in
expressions (\ref{Gaussbeta}) and (\ref{Gausss}), so, $f_\beta$ and
$f_s$ are independent of the dispersion in (\ref{Gaussmu}). We now
show that, moreover, the distributions (\ref{Gaussbeta}) and
(\ref{Gausss}) are valid not only for the Gauss distribution case,
but for all symmetric PDFs $f_{ij}$.\vspace{0.5cm}

\noindent \textit{\textbf{Statement 1.}} If
$\phi\left(q;r\right)=\phi\left(q\right)$, then
$f_s\left(y\right)=1/{\pi\sqrt{1-y^2}}$ and
$f_\beta\left(z\right)=3/{\pi\sqrt{4-z^2}}$

\noindent Indeed, it follows from (\ref{fsmu}):

\begin{equation}
f_s\left(y\right)=\frac{4}{\sqrt{3\left(1-y^2\right)}}\int_0^{+\infty}\phi\left(-t^2\right)tdt
\end{equation}

\noindent Since the integral from the right side does not depend on
$y$ and PDF $f_s\left(y\right)$ is normalized, the integral must
fail to be governed by function $\phi\left(q\right)$:

\begin{equation}\label{st1}
\int_0^{+\infty}\phi\left(-t^2\right)tdt=\frac{\sqrt{3}}{4}\left(\int_{-1}^1\dfrac{dy}{\sqrt{1-y^2}}\right)^{-1}=\frac{\sqrt{3}}{4\pi}
\end{equation}

\noindent This leads to the equation for $f_s\left(y\right)$. In the
same manner, from (\ref{fbetamu}) we easily can prove the one for
$f_\beta\left(z\right)$:

\begin{equation}
f_\beta\left(z\right)=\frac{4\sqrt{3}}{\sqrt{4-z^2}}\int_0^{+\infty}\phi\left(-t^2\right)tdt=\frac{3}{\pi\sqrt{4-z^2}}
\end{equation}\vspace{0.5cm}

\noindent The analogous notation can be made about parameter $A$ in
(\ref{PseudoGaussQR}) and distributions (\ref{PseudoGaussbeta}) and
(\ref{PseudoGausss}), which are independent of it. Let us prove that
distributions (\ref{PseudoGaussbeta}) and (\ref{PseudoGausss}) take
place not only in the pseudo-Gauss case but also for all symmetric
PDFs $f_{Q_S R_S}$:\vspace{0.5cm}

\noindent \textit{\textbf{Statement 2.}} If $f_{Q_S
R_S}\left(q;r\right)=f_{Q_S R_S}\left(q\right)$, then
$f_s\left(y\right)=1/2$ and
$f_\beta\left(z\right)=3/4\cdot\left(1-z^2\right)$

\noindent From (\ref{fsQR}) we can write:

\begin{equation}
f_s\left(y\right)=\frac{4}{3\sqrt{3}}\int_0^{+\infty}f_{Q_S R_S}\left(-t^2\right)t^4dt
\end{equation}

\noindent The same method of $f_s$ normalization provides:

\begin{equation}\label{st2}
\int_0^{+\infty}f_{Q_S R_S}\left(-t^2\right)t^4dt=\frac{3\sqrt{3}}{8}
\end{equation}

\noindent This proves the first part of the statement. Finally, from
(\ref{fbetaQR}) it turns out:

\begin{equation}
f_\beta\left(z\right)=\frac{2}{\sqrt{3}}\left(1-z^2\right)\int_0^{+\infty}f_{Q_S R_S}\left(-t^2\right)t^4dt=\frac{3}{4}\left(1-z^2\right)
\end{equation}\vspace{0.5cm}

\noindent Note that (\ref{st1}) and (\ref{st2}) are deduced by $f_s$
normalization instead of $f_{ij}$ and $f_{Q_S R_S}$ normalization.
The direct proof is more complicated and is given in Appendix 2.

\section*{4. Zeroes structure of the shuffled eigen values PDF}

\noindent Let us deduce the properties of the turbulent flow that
governs the specific shape of distributions $f_\beta$ and $f_s$,
which can be seen from figure 1 in \cite{LR} and written as:

\begin{equation}\label{conditions}
f_\beta\left(\pm1\right)=0 \;\;\;\; f^\prime_\beta\left(\pm1\right)\ne0  \;\;\;\; f_s\left(\pm1\right)\ne0 \;\;\;\; f_s\left(\pm1\right)\ne\infty
\end{equation}

\noindent Not all of these properties are independent. The
dependence of the first and the fourth properties is shown in the
article \cite{LR}. It is easy to show that the second and the third
ones are identical, too. In fact, relation (\ref{sbetas}) enables to
represent statistical properties of the flow in terms of $f_\beta$
and $f_s$, equivalently (the similar fact emerges for $f_{ij}$ and
$f_{Q_SR_S}$ expressing each other via (\ref{QRmu}), (\ref{fQRmu})
and (\ref{fmuQR})). To formulate one of the properties in a
convenient way let us introduce a secondary function $g_{ij}$ by the
following expression:

\begin{equation}\label{g}
f_{ij}\left(x;y\right)=\left|\left(x-y\right)\left(2x+y\right)\left(2y+x\right)\right|g_{ij}\left(x;y\right)
\end{equation}\vspace{0.5cm}

\noindent \textit{\textbf{Statement 3.}} If conditions (\ref{conditions}) hold, then
\begin{itemize}

\item[1)] $f_{Q_S R_S}\left(q;r\right)$ is limited on the curve $-4q^3-27r^2=0$, however, it is not identical to be equal to zero on it.

\item[2)] $g_{ij}\left(x,y\right)$ is limited on the lines $x=y$, $x=-2y$ and $2x=-y$, however, it is not identical to be equal to zero on them.

\end{itemize}
\noindent $\phi\left(q;r\right)$ can be represented uniquely in the form:

\begin{equation}\label{st3}
\phi\left(q;r\right)=\left(-4q^3-27r^2\right)^\epsilon\gamma\left(q;r\right)
\end{equation}

\noindent where $\gamma\left(q;r\right)$ is not equal to zero on the
curve $-4q^3-27r^2=0$ and does not tend to infinity on it. By
substituting the latter expression in (\ref{fsmu}) and taking into
consideration relations (\ref{conditions}) one can get
$\epsilon=1/2$. Applying conditions (\ref{fQRmu}) and (\ref{fmuQR}),
we finally prove the statement.\vspace{0.5cm}

\noindent Therefore, shuffled eigenvalues PDF, which has the
Gaussian shape in case of Gauss normal velocity distribution, in
fact, vanishes on the lines $x=y$, $x=-2y$ and $2x=-y$ corresponding
to axisymmetric extension $\beta=+1$ (or $s=+1$) and axisymmetric
contraction $\beta=-1$ (or $s=-1$).

\section*{5. Conclusions}

\textit{Comparison of the article results with the previous ones}\vspace{0.5cm}

\noindent Let us start our conclusions with considering some
important difference between the results of our article and the
results of the article \cite{LR}, which are believed to be right in
a wide range of articles following \cite{LR} (see the second
footnote).

\noindent The key distinction is that we deduce PDF $f_\beta$ of the
dimensionless parameter $\beta$ (\ref{beta}) (in section 3) to be
nonzero in the ends of the function domain in the Gauss distribution
case. Likewise, PDF $f_s$ of parameter $s$ (\ref{s}) in our article
is deduced (in section 3) to tend to infinity in the ends of its
domain, but in \cite{LR} it was found to be constant on its domain
$\left(f_s=1/2\right)$. The fact that $f_s$ is uniform for the Gauss
distribution was believed to be the reason for choosing $f_s$ as the
most convenient single PDF to illustrate the tendencies of the fluid
particles strain in the turbulent flow \cite{Meneveau, LR}.
Following this logic it should be mentioned that $\beta$ PDF, which
is not constant in the case of the Gauss velocity distribution (see
Fig. 2) is more preferable, because it is quite close to be constant
(it can possess values from only $f_\beta\left(0\right)\simeq0.48$
to $f_\beta\left(\pm1\right)\simeq0.55$). As for $s$ PDF, it tends
to infinity at the ends of its domain in the Gauss case (see Fig. 2)
and, thus, is far from being constant.

\noindent From the physical point of view this agrees with the
article \cite{Ashurst} conclusions that the axisymmetric extension
and contraction of fluid particles ($\beta=\pm1$ or $s=\pm1$) do not
exist in the turbulent flow. It should be mentioned that the absence
of $\beta=\pm1$ ($s=\pm1$) particles surely does not mean the
prevalence of filament structures because the whole interval of
values $0<\beta\le1$ leads to the pancake shape and as can be seen
from the experimental and numerical results for $f_\beta$ (e.g.
figure 1 in \cite{LR}):

\begin{equation}\
\int_0^1f_\beta\left(z\right)dz>\int_{-1}^0f_\beta\left(z\right)dz
\end{equation}\vspace{1cm}

\noindent\textit{Pseudo-Gauss distribution and statement 3}\vspace{0.5cm}

\noindent In the present article we found the class of distributions
(statement 2 in section 3), for which $f_\beta$ and $f_s$ have the
form that they seemed to have in case of the Gauss distribution (see
\cite{LR} and the articles from the second footnote on page 3). We
called the exponential distribution from this class
(\ref{PseudoGaussQR}) the pseudo-Gauss distribution and it differs
from the Gauss distribution in the absence of the pre-exponential
factor $\left(-4q^3-27r^2\right)^{-1/2}$ , which, as we concluded,
was not accounted before. This pre-exponential factor leads to the
vanish of pseudo-Gauss PDF $f_{ij}$ (\ref{PseudoGaussmu}) on the
lines corresponding to axisymmetric extension and contraction. Thus,
it dramatically differs from the Gauss distribution, which vanishes
nowhere. We saw in section 4 that such structure is the result of
vanishing $f_\beta$ at $\pm1$ and, thus, the latter property is not
a trivial one (i.e. kinematical in terms of \cite{LR, Shtilman}),
but an essential property of the turbulent statistics (i.e.
dynamical).

\noindent Let us pay attention to the fact that has a physical
meaning. An energy cascade from large to small scales is known to
arise in the turbulent flow \cite{Frisch} that means macroscopic
irreversibility of the turbulent processes. One can conclude that
the irreversibility of fluid dynamics appear due to skewness of its
PDF as a result of the prevalence of direct processes over the
inverse ones. For instance, the Gauss velocity distribution
(\ref{Gaussmu}) is symmetric and, thus, free of the irreversibility.
This is believed to be the keynote distinction between the Gauss and
real turbulent velocity distributions. Similarly, via its symmetry,
the pseudo-Gauss distribution (\ref{PseudoGaussmu}) corresponds to
the reversible process and does not contain the irreversible energy
cascade. We showed in section 3 that the pseudo-Gauss distribution
is qualitatively more suitable for the real turbulent statistics and
we generalized this fact in statement 3 (in section 4) on any
$f_{ij}$, deducing $f_{ij}$ to vanish on these lines to satisfy
conditions (\ref{conditions}). Hence, we find a new ("reversible")
reason for the turbulent velocity field nongaussianity.

\noindent However, statement 3 is not deduced from the first
principles. In fact, it is the consequence of experimental and DNS
results (\ref{conditions}). Although there are no significant doubts
that such a great number of the concordant data gives real
information about turbulent statistics, the fundamental reasons and
consequences of the unexpected nongaussianity need to be explored in
the future.

\noindent The first step in this way can be done by finding the
example of reversible velocity field statistics, which governs the
pseudo-Gauss distribution or any distribution from the class
concerned in statement 2. Unfortunately, we have not found it
yet.\vspace{0.5cm}

\noindent\textit{Joint PDFs of the shuffled eigen values and integral transformations}\vspace{0.5cm}

\noindent We think that the PDF of the strain-rate tensor eigen
values $f_{ij}$, which was introduced in section 2, will be
convenient for the further research of the tensor statistical
properties (especially, in their theoretical investigations) due to
its unbounded domain and symmetric properties.

\noindent The transformations (\ref{fQRmu})-(\ref{fbetamu}) that we
found in section 2 by means of $f_{ij}$ were useful in deducing the
article statements. We believe that these transformations will be
useful not only in future theoretical investigations but also for
the validation of numerical calculations and experimental results.
We also think that the method of finding the exact transformations
could give many fruitful results about the velocity gradient tensor,
too. In fact, there are many other dimensionless parameters
depending also on the vorticity vector (e.g see \cite{Meneveau, Xu2,
Wallace, Chamecki}) and one may hope that deduction of analogous
transformations for them could give further nontrivial conclusions
about turbulent velocity gradient statistics.

\section*{6. Appendix}

\noindent\textit{1. Ordered eigen values and their joint PDF}\vspace{0.5cm}

\noindent There are three pairs of independent ordered eigenvalues.
They can be chosen as $\lambda_1$ and $\lambda_2$ ; $\lambda_1$ and
$\lambda_3$ ; $\lambda_2$ and $\lambda_3$. Thus, there are six
different joint PDFs of ordered eigen values: $f_{\lambda_1
\lambda_2}$ and $f_{\lambda_2 \lambda_1}$, $f_{\lambda_1 \lambda_3}$
and $f_{\lambda_3 \lambda_1}$, $f_{\lambda_2 \lambda_3}$ and
$f_{\lambda_3 \lambda_2}$. The functions domains given in Fig. 4
have intersections only on lines $x=y$, $x=-x-y$ and $y=-x-y$, that
correspond to axisymmetric extension or contraction.

\begin{figure}[h]
\center{\includegraphics[scale=0.75]{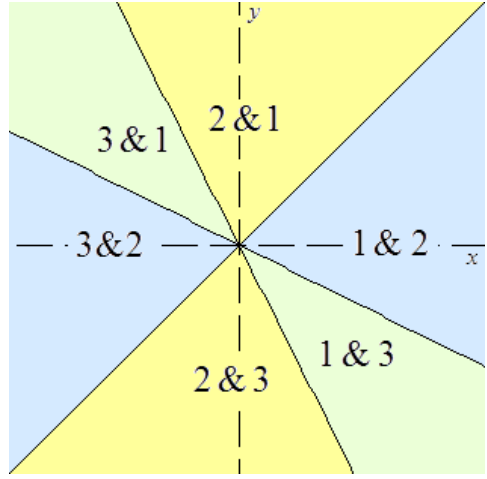}} \caption{\small{The
ordered eigen values PDF domains: $i\&j$ of $\lambda_i$ and
$\lambda_j$.}}
\end{figure}

\noindent Due to statement 3 it is those lines, on which $f_{ij}$ is
equal to zero. The relations between these functions and shuffled
eigenvalues PDF are very simple and PDFs of any values $\lambda_i$
and $\lambda_j$ $\left(i\ne j\right)$ can be expressed (in their
domain!) by the fully identical formulas:

\begin{equation}
f_{\lambda_i \lambda_j}\left(x;y\right)=6f_{ij}\left(x;y\right)
\end{equation}

\noindent Hence, by virtue of statement 3 each function
$f_{\lambda_i \lambda_j}\left(x;y\right)$ vanishes on the bound of
its domain. Despite the simplicity of the latter relation, we
consider our PDF of shuffled eigenvalues more useful, especially in
theoretical research, since it has an unbounded domain and symmetric
properties.\vspace{0.5cm}

\noindent\textit{Another deduction of the statement 1.}\vspace{0.5cm}

\noindent Since the deductions of statements 1 and 2 are almost
similar we give direct deduction of statement 2 only. By condition
of statement 2 we have:

\begin{equation}
f_{Q_S R_S}\left(q;r\right)=f_{Q_S R_S}\left(q\right)
\end{equation}

\noindent It provides from the normalization of $f_{Q_S R_S}$ that:

\begin{multline}
1=\iint_{D_{Q_S R_S}} f_{Q_S R_S}
\left(q\right)dqdr=\\=\int_{-\infty}^0 f_{Q_S R_S} \left(q\right)
dq\int_{-\frac{2}{3\sqrt{3}}\left(-q\right)^{3/2}}^{\frac{2}{3\sqrt{3}}\left(-q\right)^{3/2}}dr=\frac{4}{3\sqrt{3}}\int_{-\infty}^0\left(-q\right)^{3/2}f_{Q_S
R_S}\left(q\right)dq
\end{multline}

\noindent Finally, using (\ref{fsQR}) we have:

\begin{multline}
f_s\left(y\right)=\frac{4}{3\sqrt{3}}\int_{0}^{+\infty}f_{Q_S
R_S}\left(-t^2\right)t^4dt=\\=\frac{4}{3\sqrt{3}}\int_{-\infty}^0\left(-q\right)^2f_{Q_S
R_S}\left(q\right)\frac{dq}{2\left(-q\right)^1/2}=\frac{1}{2}
\end{multline}

\noindent Statement 1 can be proved identically using (\ref{fQRmu}), (\ref{fsmu}), (\ref{fbetamu}).

\end{document}